\documentclass[superscriptaddress,twoside,twocolumn,bibnotes,floats,twoside,aps]{revtex4}
\usepackage{amsmath}
\usepackage{graphicx}
\usepackage{times}
\graphicspath{{../Figures/}{.}}
\begin{document}
\title{\bf Stratified spatiotemporal chaos in anisotropic reaction-diffusion 
systems}

\affiliation{Max-Planck-Institut f\"ur Physik komplexer Systeme,
N\"{o}thnitzer Str. 38, 01187 Dresden, Germany}
\affiliation{
Center for Nonlinear Studies and T-7,
Theoretical Division, Los Alamos National Laboratory, 
Los Alamos, NM 87545
}
\affiliation{
The Jacob Blaustein Institute for Desert Research 
and the Physics Department,
Ben-Gurion University, Sede Boker Campus 84990, Israel
}

\author{Markus B\"ar}
\email{baer@idefix.mpipks-dresden}
\affiliation{Max-Planck-Institut f\"ur Physik komplexer Systeme,
N\"{o}thnitzer Str. 38, 01187 Dresden, Germany}
\affiliation{
The Jacob Blaustein Institute for Desert Research 
and the Physics Department,
Ben-Gurion University, Sede Boker Campus 84990, Israel
}
\author{Aric Hagberg}
\email{aric@lanl.gov} 
\homepage{http://math.lanl.gov/~aric}
\affiliation{
Center for Nonlinear Studies and T-7,
Theoretical Division, Los Alamos National Laboratory, 
Los Alamos, NM 87545
}

\author{Ehud Meron}
\email{ehud@bgumail.bgu.ac.il}
\affiliation{
The Jacob Blaustein Institute for Desert Research 
and the Physics Department,
Ben-Gurion University, Sede Boker Campus 84990, Israel
}

\author{Uwe Thiele}
\email{thiele@nolineal.pluri.ucm.es}
\affiliation{Max-Planck-Institut f\"ur Physik komplexer Systeme,
N\"{o}thnitzer Str. 38, 01187 Dresden, Germany}
\affiliation{Instituto Pluridisciplinar, Universidad Complutense Madrid, 
    Paseo Juan XXIII  1, E-28040 Madrid, Spain}

\received{Submitted: 6 April 1999; }
\revised{Revised: 28 July 1999}

\begin{abstract}
Numerical simulations of two dimensional pattern formation 
in an anisotropic bistable
reaction-diffusion medium reveal a new dynamical state, 
stratified spatiotemporal chaos, characterized by strong correlations
along one of the principal axes. Equations that describe 
the dependence of front motion on the angle illustrate 
the mechanism leading to stratified chaos.
\end{abstract} 

\pacs{PACS numbers:5.45; 5.70; 82.65 }
\maketitle

Pattern formation in nonequilibrium systems has been extensively
studied in isotropic, two-dimensional media~\cite{CroHo}. 
Among the most prominent experimental examples are 
Rayleigh-Benard convection and the Belousov-Zhabotinsky reaction 
in the contexts of fluid dynamics and chemical reactions respectively. 
Recently, there has also been considerable interest 
in systems with broken rotational 
symmetry, like convection in liquid crystals~\cite{KraPe} 
and chemical waves in catalytic surface reactions~\cite{ImbErtl}. 
Experimental and theoretical studies of such anisotropic systems showed 
novel phenomena like ordered arrays of topological defects~\cite{chevrons},
anisotropic phase turbulence~\cite{Faller},
reaction-diffusion waves with sharp corners~\cite{corners}, 
and traveling wave fragments along a preferred orientation~\cite{MertPRE}. 
Anisotropy is also often present in pattern formation processes 
in biological media, {\it e.g.} in cardiac tissue~\cite{cardiac}.

In this Letter we present a new dynamical state that is possible only
in anisotropic media - stratified spatiotemporal chaos.  We
demonstrate the phenomenon with numerical simulations of the bistable
FitzHugh-Nagumo equations with anisotropic diffusion and
characterize it by computing orientation dependent correlation
functions.  In addition, an equation for the dependence of front
velocities on parameters, curvature, and the angular
orientation is derived. The mechanism leading to stratified chaos
is described in terms of these analytic results. 
It is tightly linked to the anisotropy of the system and  
differs from the mechanism 
leading to spiral chaos in isotropic bistable
media~\cite{LeePRE,HaMe94,HaMe97}.
The findings here are relevant to catalytic reactions on surfaces
where anisotropy is naturally provided by crystal symmetry and in
biological tissue where anisotropy comes from fiber orientation.

    
Many qualitative features of pattern formation in chemical and
biological reaction-diffusion systems are well described by
FitzHugh-Nagumo (FHN) type models for bistable
media~\cite{IMN89,HaMeNL}.  The specific model we choose to study is
\begin{eqnarray}
\frac{\partial u}{\partial t} &=&\epsilon ^{-1}(u-u^{3}-v)+\delta
^{-1}\nabla ^{2}u+\frac{\partial }{\partial y}\bigl[d\delta
^{-1}\frac{ \partial u}{\partial y}\bigr]\,, \nonumber \\
\frac{\partial v}{\partial t} &=&u-a_{1}v-a_{0}+\nabla ^{2}v\,,
\label{afhn}
\end{eqnarray}
where $u$ is the activator and $v$ the inhibitor.  The parameters
$a_{1}$ and $a_0$ are chosen so that Eqs.~(\ref{afhn}) represent a
bistable medium with two stationary and uniform stable states, an
``up'' state, $(u_{+},v_{+})$, and a ``down'' state, $(u_{-},v_{-})$.
Front solutions connect the two states.  
The number of front solutions changes, when a single front (an ``Ising''
front) that exists for values of
$\eta := \sqrt{\epsilon\delta} > \eta_{c}$ loses stability to a pair
of counter-propagating fronts (``Bloch'' fronts) for $\eta \le \eta
_{c}$.  The corresponding bifurcation is referred to as the nonequilibrium 
Ising-Bloch (NIB) bifurcation, hereafter the ``front bifurcation.''
The anisotropy of the medium is reflected through the parameter $d$.  

\begin{figure}[htb]
\centering\includegraphics[width=2.5in]{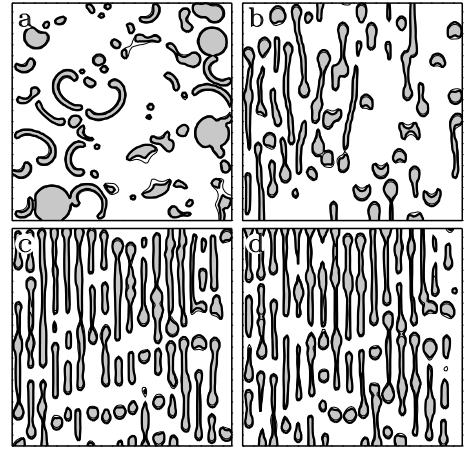}
\caption{
The development of stratified spatio-temporal chaos  from an initial
state of isotropic spatio-temporal chaos.  The thick contour line
represents $u=0$ and the thin contour line $v=0$.  The shaded
regions are up-state domains. 
Parameters: $a1=2.0$, $a_0=-0.1$, $\epsilon=0.039$, $\delta=1.7$,
$d=1.0$, $x=[0,80]$, $y=[0,80]$.
}
\label{fig:chaos}
\end{figure}

Figure~\ref{fig:chaos} shows the formation and time evolution of 
stratified chaos obtained by numerical solution of
 Eqs.~(\ref{afhn}). The initial state
is isotropic spiral chaos. As time evolves a clear orientation of up-state
(grey) domains along the $y$ direction develops. The domains consist 
primarily of elongating stripe segments which either merge with other 
segments or shorten by emitting traveling blobs. 
The stratified chaos state is robust and develops from a variety of initial 
conditions including a single spot.

\begin{figure}[htb] 
\centering\includegraphics[width=2.5in]{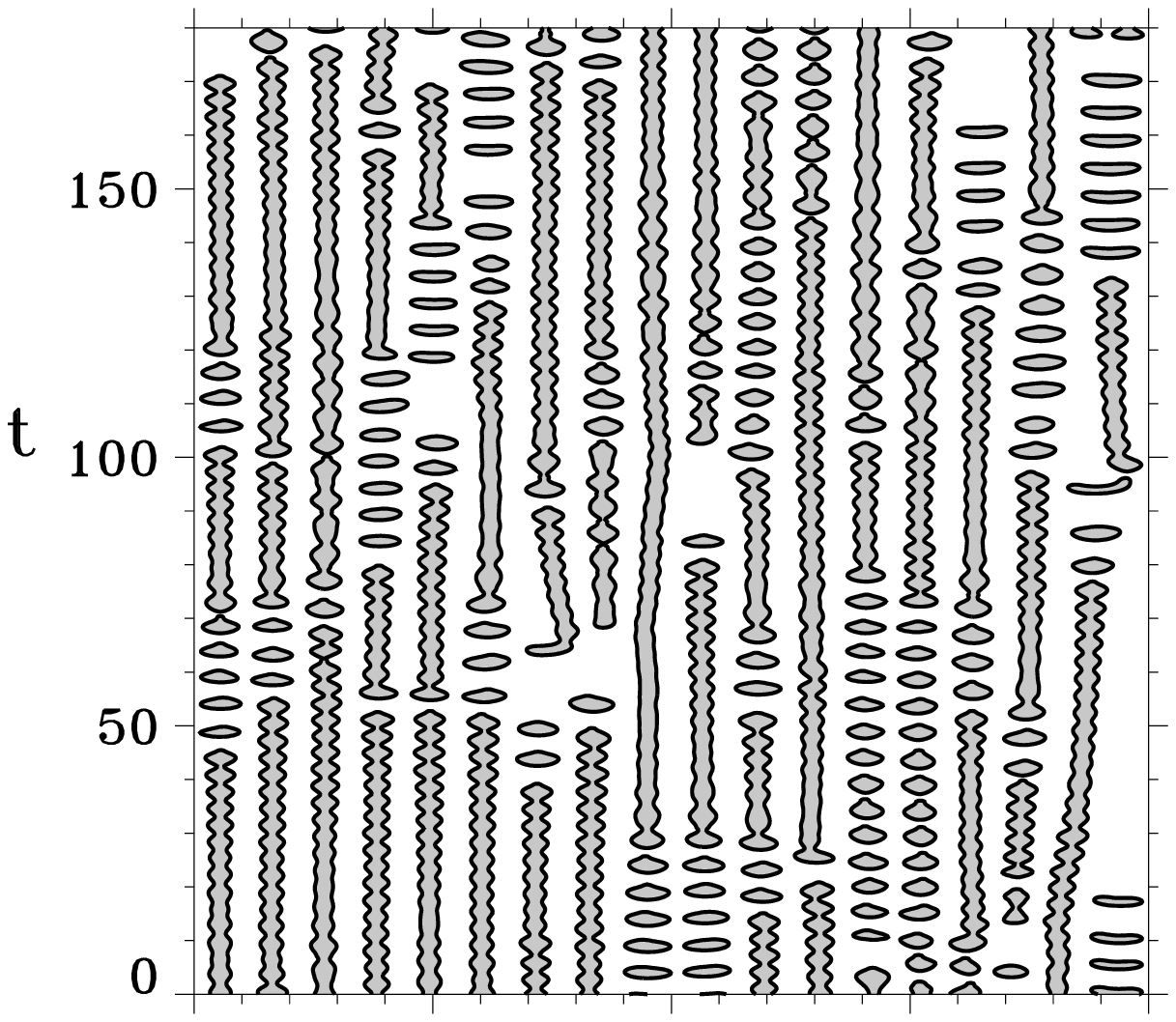}
\centering\includegraphics[width=2.5in]{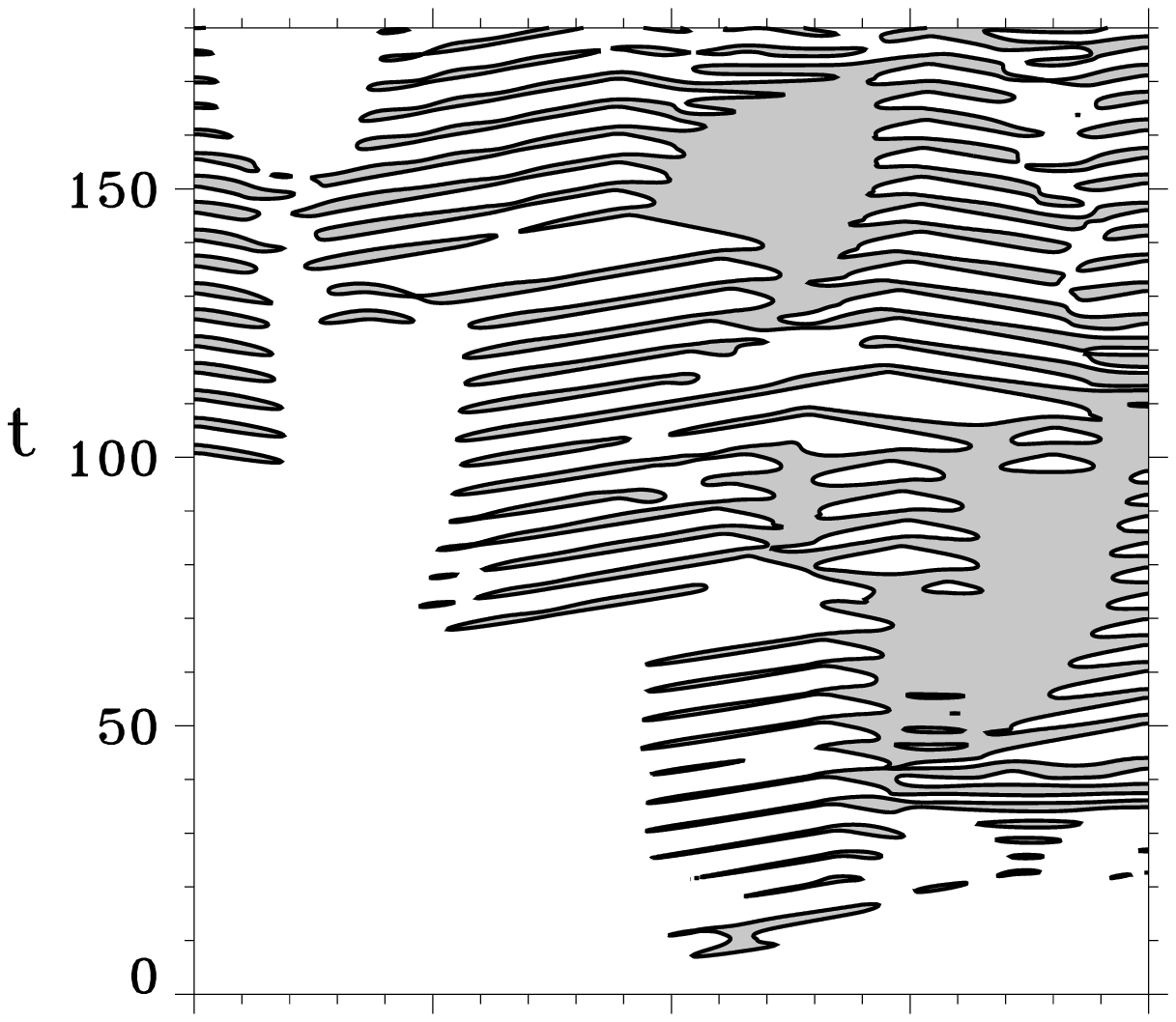}
\caption{Space-time plots of cuts parallel to the $x$ axis (top) and  
the $y$ axis (bottom) for the simulation of
Fig.~\protect\ref{fig:chaos}. 
The nearly vertical columns on the top figure show instances of 
periodic breathing motion of stripe segments (modulated continuous segments)
and periodic blob formation (spot arrays). 
}
\label{fig:spacetime}
\end{figure}

\begin{figure}[htb] 
\centering\includegraphics[width=3.0in]{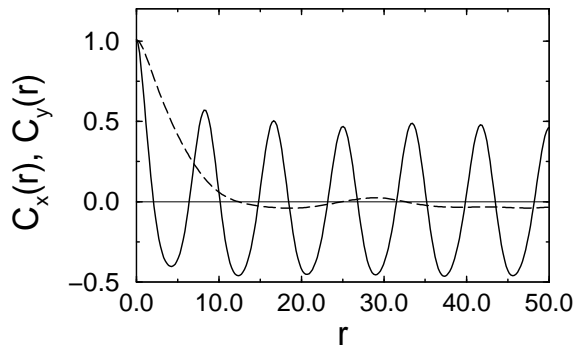}
\caption{ Correlation functions $C_x(r)$ (solid curve) and 
$C_y(r)$ (dashed curve).  
}
\label{fig:correlation}
\end{figure}

To gain insight we have followed the dynamics of $u$ and $v$
along the $x$ and $y$ axes and display them in the form of
space-time plots in Fig.~\ref{fig:spacetime}.  A nearly periodic
non-propagating pattern along the $x$ axis and irregular
traveling wave phenomena along the $y$ axis are observed.  
A characteristic property 
of spatio-temporal chaotic patterns are correlations that decay 
on a length scale $\xi$
much smaller than the system length $L$.  We have computed the normalized 
spatial two-point correlation functions, $C_y(r)$ and $C_x(r)$, for the $u$ 
field in both the $x$ and $y$ directions, where 
$C_y(r) = < \Delta u(x,y+r) \Delta u(x,y) > / <\Delta u(x,y)^2>$, 
$C_x(r) = < \Delta u(x+r,y) \Delta u(x,y) >/ <\Delta u(x,y)^2> $,
$\Delta u(x,y) = u(x,y) - <u>$, and the brackets $<>$
denote space and time averaging. Figure~\ref{fig:correlation} shows the 
results of these computations.  Correlations in the $y$
direction decay fast to zero, whereas correlations in the $x$ direction 
oscillate with constant amplitude.  This observation may be used to define 
stratified chaos as a state that displays finite correlation length in one 
direction ($x$) and infinite correlation length in the other ($y$).

An important analytical tool for studying front dynamics consists of 
relations between the normal front velocity $C_n$ and other front 
properties like curvature. 
Relations of that kind have successfully been used in the study
of pattern formation in isotropic media~\cite{HaMe94},
and we wish to exploit this tool for anisotropic media as well.

Velocity-curvature relations are derived here for 
$\lambda:=\sqrt{\epsilon/\delta}\ll 1$.  The first step in this derivation
is to define an orthogonal
coordinate system $(r,s)$ that moves with the front, where $r$ is a
coordinate normal to the front and $s$ is the arclength. We denote the
position vector of the front by ${\bf X}(s,t)=(X,Y)$, and define it to
coincide with the $u=0$ contour line.  The unit vectors tangent and
normal to the front are given by
${\bf \hat s}=\cos\theta{\bf \hat x} + \sin\theta{\bf \hat y}$
and ${\bf \hat r}=-\sin\theta{\bf \hat x} + \cos\theta{\bf \hat y}$,
respectively, where $\theta(s,t)$ is the angle that $\hat s$ makes with the $x$
axis. A point ${\bf x}=(x,y)$ in the laboratory frame can be expressed
as
${\bf x}={\bf X}(s,t)+r{\bf \hat r}$.
This gives the following relations between the laboratory coordinates
$(x,y)$ and the coordinates $(s,r)$ in the moving frame:
$x=X(s,t)-r\sin\theta(s,t)$, 
and $y=Y(s,t)+r\cos\theta(s,t)$
where we defined ${\bf \hat s}=\partial{\bf X}/\partial s$ and ${%
\partial X}/{\partial s}=\cos\theta$, ${\partial Y}/{\partial
s}=\sin\theta$.  In terms of the moving frame coordinates the front normal
velocity and curvature are given by
$C_{n}=-{\frac{\partial r}{\partial t}}$ and 
$\kappa =-\frac{\partial \theta }{\partial s}$, respectively.

The second step is to express Eqs.~(\ref{afhn}) in the moving frame and use 
singular perturbation theory, exploiting the smallness of $\lambda$.  
We distinguish between an inner region that includes the narrow front 
structure, and outer regions on both sides of the front. In the inner region
$\partial u/\partial r\sim {\cal O}(\lambda ^{-1})$ and 
$\partial v/\partial r\sim {\cal O}(1)$. In the outer regions both 
$\partial u/\partial r$ and $\partial v/\partial r$ are of order unity. 
In the inner region $v=v_f$ is taken to be constant. Expanding both $u$ and
$v_f$ as powers series in $\lambda$ and using these expansions in the moving 
frame equations we obtain, at order 
${\cal O}(\lambda)$, a solvability condition that leads to the equation 
\begin{equation}
C_{n}=-{\frac{3}{\eta \sqrt{2}}}I(\theta )v_{f}-\frac{1+d}{\delta I(\theta
)^{2}}\kappa \,,  \label{Cn}
\end{equation}
where $I(\theta)=\sqrt{1+d\cos^2\theta}$.
In the outer regions to the left and to the right of the 
front region different approximations can be made. Here 
$\frac{\partial u}{\partial r}\sim\frac{\partial v}{\partial r}
\sim{\cal O}(1)$ and to leading order all terms that contain the factor 
$\lambda$ can be neglected. The resulting equations can be solved for $v$
in the two outer regions. Continuity of $v$ and of 
$\frac{\partial v}{\partial r}$
at the front position $r=0$ yield a second relation between $C_n$ and $v_f$.
Eliminating $v_f$ by inserting this relation into Eq.~(\ref{Cn}) gives an
implicit relation between the normal velocity of the front and its
curvature 
\begin{equation}
 C_n+\frac{1+d}{\delta I(\theta)^{2}}\kappa= 
 \frac{3I(\theta)(C_n+\kappa)}{\eta\sqrt{2}q^2\sqrt{(C_n+\kappa)^2+4q^2}} 
+\frac{3I(\theta)a_0}{\eta\sqrt{2}q^2}\,,  \label{CnK}
\end{equation}
where $q^2 = a_1 + 1/2$. A complete account of this derivation will be 
published elsewhere. 

\begin{figure}[htb]
\centering\includegraphics[width=2.8in]{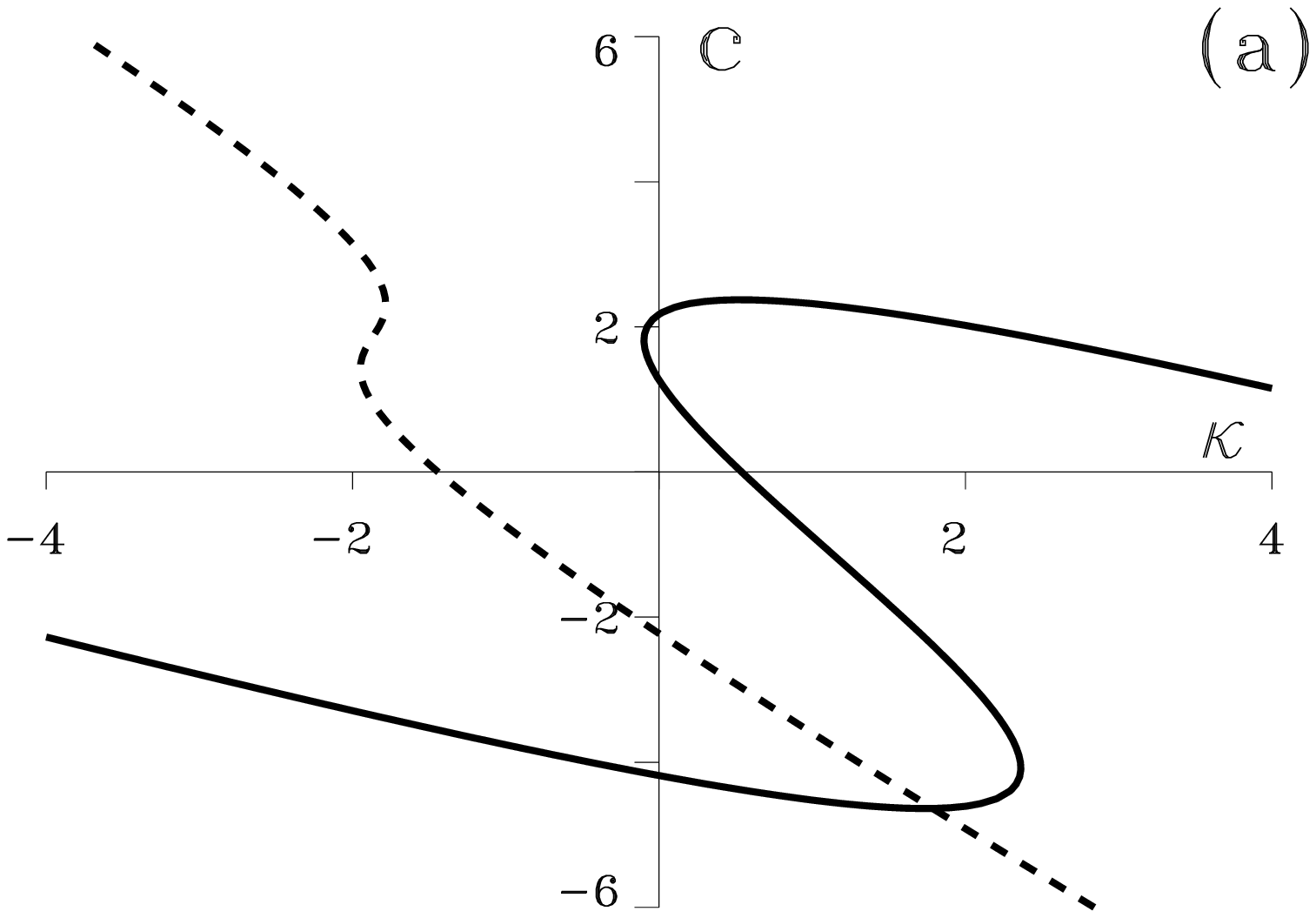}
\centering\includegraphics[width=2.8in]{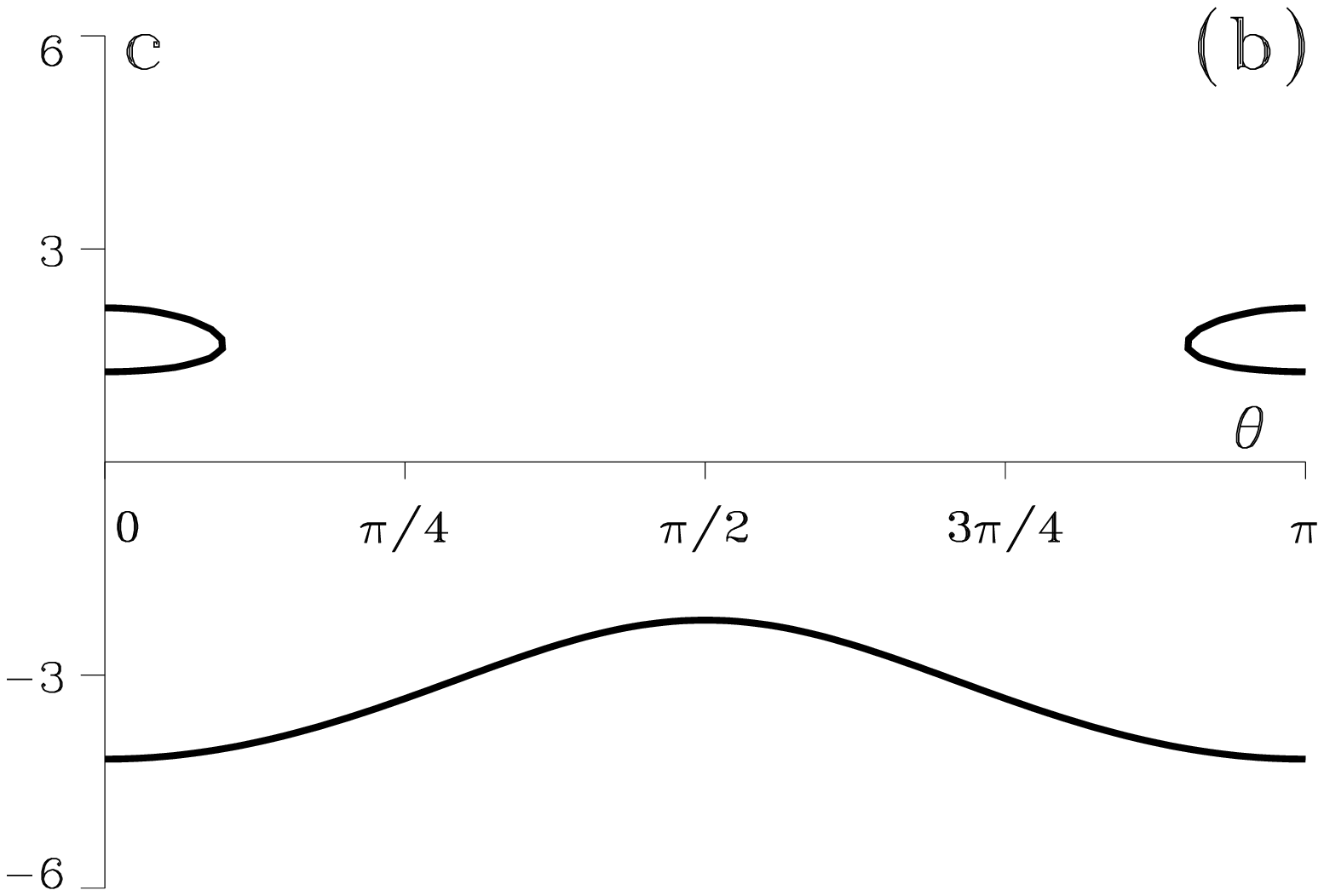}
\caption{Two views of the relation ~(\protect\ref{CnK})
corresponding to the parameters of the simulation in Figure 1.
(a) The velocity-curvature relation for fronts in the 
$x$ direction (dashed curve) and $y$ direction (solid curve). 
(b) The velocity of planar ($\kappa=0$) fronts at different angles.
Bloch fronts exist in narrow sectors around $\theta=0$ and $\theta=\pi$.
The wider sectors in between correspond to Ising fronts.}
\label{fig:cvk}
\end{figure}

Figures ~\ref{fig:cvk} display solutions of Eq.~(\ref{CnK}) showing 
the dependence of the front velocity on front curvature and
propagation direction for the parameter values of 
Fig.~\ref{fig:chaos}.  In Fig.~\ref{fig:cvk}a $C_n$ vs $\kappa$ relations are 
shown for two orthogonal propagation directions. In the $x$ direction,
$\theta=\pi/2$ (dashed curve), there is only one planar front solution
with negative velocity (a down-state invading an up-state),  
indicating an Ising regime. The negative slope of the $C_n$ $\kappa$ relation
implies stability to 
transverse perturbations. In the $y$ direction, $\theta=0$ (solid curve), 
there are three planar front solutions indicating a Bloch regime. The 
positive-velocity front (up-state invading down-state) is unstable to 
transverse perturbations whereas the negative-velocity front is stable.
The middle branch corresponds to an unstable front.

Figure~\ref{fig:cvk}b shows the angular dependence of planar-front velocities.
Counter-propagating (Bloch) fronts exist in a narrow sector around the $y$ 
direction ($\theta=0$).
The other directions support only a single (Ising) front.
The Ising
front speed is smallest in the $x$ direction ($\theta=\pi/2$) and increases
as $\theta$ deviates from $\pi/2$. 

The dynamics of fronts as displayed in Fig.~\ref{fig:chaos} are affected
by curvature, propagation direction, and front interactions.
Eq.~(\ref{CnK}) captures the effects of the first two
factors but does not contain information about front interactions. The 
necessary information for our purpose can be summarized as follows. 

The time evolution of a pair of fronts approaching one another is
affected by the speed of non-interacting fronts and by the diffusion
rate of the activator $u$ (assuming an inhibitor diffusion constant
equal to unity as in Eqs.~(\ref{afhn})).  Consider a pair of fronts
pertaining to up-states invading a down-state, propagating toward one
another in an isotropic medium ($d=0$).  If the distance between the
fronts decreases below a critical value, $\lambda_c\sim {\cal
O}(\sqrt{\epsilon/\delta})$, the two fronts collapse, leaving a
uniform up-state. The accumulation of the inhibitor in the space
enclosed by the fronts, however, slows their motion.  If the
(non-interacting) front speed is low enough, or if $\lambda$ is small
enough, there is enough time for the inhibitor to slow the fronts down
to a complete stop before they reach the critical distance
$\lambda_c$. In that case the subsequent evolution depends on the
front type.  A pair of Ising fronts may either form a stationary pulse
or, closer to the front bifurcation, a breathing pulse.  A pair of
Bloch fronts reflect and propagate away from one
another~\cite{HaMeNL}.  Thus, high front speed or fast activator
diffusion ($\delta$ small and $\lambda_c$ large) lead to collapse,
whereas low front speed and slow activator diffusion lead to strong
front repulsion.  A similar argument holds for down-state invading
up-state fronts.

Returning to anisotropic media we need to know how the two factors
that affect front interactions, front speed and activator diffusion,
depend on the direction of propagation.  The angular dependence of the
front speed is already given in Fig.~\ref{fig:cvk}b. The angular
dependence of the activator diffusion can be deduced by inspecting
Eqs.~(\ref{afhn}). It is $1/\delta$ in the $x$ direction and
$(1+d)/\delta$ in the $y$ direction. Since $d>0$ the activator
diffusion constant increases as the propagation direction changes from
the $x$ direction to the $y$ direction.

We can discuss now the mechanism of stratified chaos. As
Fig.~\ref{fig:cvk}a shows, the $x$ direction represents a system 
that supports stationary or breathing planar stripe
patterns. The $y$ direction represents a system 
where traveling waves prevail.  The distinct characters of the medium
along the two principal axes is reflected in the space-time plots of
Fig.~\ref{fig:spacetime}: nearly vertical columns in the $x$ direction
indicate stationary or breathing motion, and diagonal stripes in
the $y$ direction indicate traveling wave phenomena.

\begin{figure}[htb]
\centering\includegraphics[width=3.0in]{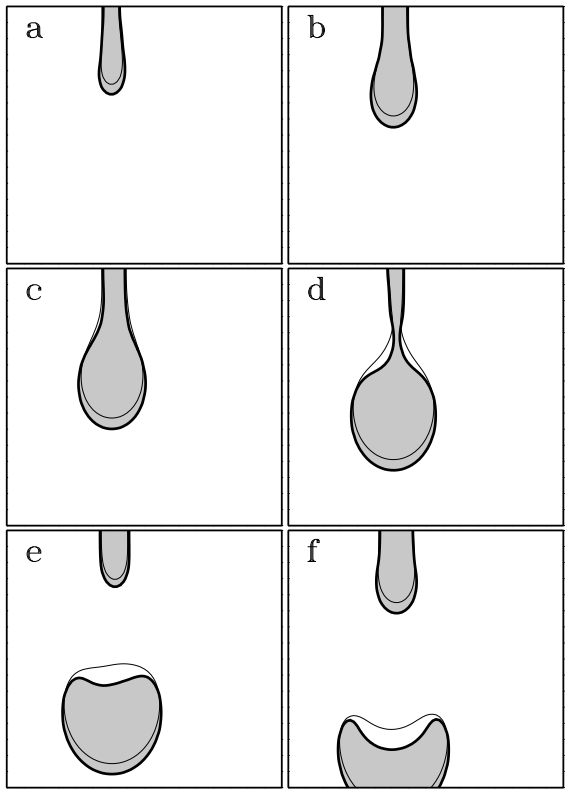}
\caption{Close up of repeated blob formation. Shaded regions are up-state
domains. Thick (thin) lines are $u=0$ ($v=0$) contours. The $v=0$ line
always lags behind the $u=0$ line.  The tip of a stripe segment (a)
grows outward (b)-(c). A pinching dynamics begins (d) which leads to
blob formation (e) traveling along the $y$ direction (f). The blob
formation leaves a shortened stripe segment (e) whose tip grows
outward again (f) and the process repeats.  The parameters are the
same as in Fig.~\protect\ref{fig:chaos}.  }
\label{fig:blob}
\end{figure}

The irregular character of the dynamics comes from blob formation
events as shown in Figure~\ref{fig:blob}. The front speed and
activator diffusion in the $x$ direction are sufficiently slow for
Ising fronts to repel one another (rather than collapse).  As the tip
of the stripe segment grows outward and forms a bulge, propagation
directions deviating from the $x$ direction develop. At these
directions both the front speed and the activator diffusion are
higher.  As a result approaching fronts may collapse. This is exactly
the blob pinching process in frames (c), (d) and (e) of
Fig.~\ref{fig:blob}. The process is periodic for extended periods of
time as indicated by the vertical spot arrays in
Fig.~\ref{fig:spacetime}.


In summary, stratified chaos relies on two main elements: (i) stationary
or breathing domains vs traveling wave phenomena in orthogonal
directions, and (ii) an angular dependence of front interactions that leads to
blob formation.  Without the second element stripe segments would
merge to ever longer segments until a periodic stripe pattern is
formed.  These elements suggest the parameter regime where stratified
chaos is expected to be found. The first element implies a regime
along the front bifurcation such that there is stationary or breathing
motion in the direction with faster activator diffusion and 
traveling waves in the other
(slower diffusion) direction.  The width of this regime increases with
the anisotropy $d$. The second element implies that in the direction
of breathing or stationary domains the system is close to the onset of
breathing motion. Deviations from this direction which move the system
toward the traveling wave regime then lead to front collapse.  These
expectations were verified numerically.  Part of the work of U.~T. was
supported by grant D/98/14745 of the German Academic Exchange Board
(DAAD).  M.~B. gratefully acknowledges the hospitality of Ben-Gurion
University and support of the Max-Planck-Society (MPG) by an
Otto-Hahn-Fellowship.  Part of this research is 
supported by the Department of Energy, under contract W-7405-ENG-36.

\end{document}